# Computer Simulations on Barkhausen effects and Magnetizations in Fe Nano-Structure Systems

Shuji Obata*

*School of Science & Engineering, Tokyo Denki University,*

*Ishizaka, Hatoyama, Hiki, Saitama 350-0394, Japan*

(                           )

The magnetization processes in Fe nano-systems are investigated using the numerical simulations based on classical magnetic dipole moment interactions. The domain energies are calculated from moment-moment interactions over whole systems using large scale computing resources. The results directly show most of basic magnetization phenomena. The Barkhausen effects are represented with jumps and terraces of magnetization steps induced from external field changes of $\Delta H$.

*obata@mail.dendai.ac.jp

## 1. Introduction

In recent studies, it has been cleared that characters of nano-scale magnetic materials show completely different from those in bulk systems, which are determined from the boundary conditions and the body shapes. As for Fe, the nano-scale systems induce strong coercivities $H_c$ and high remanent magnetizations $B_r$ as the hard magnetic materials, where the bulk systems only induce the soft magnetic characteristics. The bulk systems constructed from these nanostructured local compositions show the various particular characteristics. These scientific problems have been investigated as the Bulk Nano-Structured Materials in recent years. For clearing such phenomena, in this paper, the magnetization curves in Fe nano-scale systems are investigated based on long range classical magnetic dipole moment interactions as a theoretical study. Realistic magnetizations are cleared by the use of the large scale computing resources, where exact results have not been produced till in these days.[1-5] Nevertheless, the correct theory already existed, the first stage studies of these in a century ago could not correctly discuss the magnetizations because of very few computing resources.

A simple cubic structure system of Fe has no magnetic phenomenon in one domain. It needs some type anti-magnetized walls to make ferromagnetic states. These results are obtained through large scale computing in large systems. Generally, magnetic characters in industrial devices are distinguished to soft or hard magnetization, which phenomena are induced from the magnetic domain structures in large scale systems. In a recent study by Koyama et al,[3] the hysteresis curves of FePt are calculated using a Phase-Field method including the long range dipole moment interactions, where full interactions in a whole system are considered, and time delay of magnetizations is taken into account. Precise *B-H* curves could be theoretically calculated only using such *long range interactions* including the *time delay magnetizations*. But, it might be impossible to seek out the reference explaining clearly such two key words now.

The magnetizations and the Barkhausen (B) noises are explained using the domain energy systems equated in §2.1, and the basic energy factors between two moments are formulated in §2.2. The nano-scale simulations using the classical magnetic dipole moment interactions are performed in §3 for representing the character differences in nao-systems caused by only differences of the body shapes. The structures of nano-Fe systems are explained in §3.1. Four numerical simulations are executed about a thin film structure in §3.2, a cubic structure in §3.3, a short belt structure in §3.4 and a long belt structure in §3.5 respectively. The B effects are realized as transitions of domain structures along with changes of flow out fluxes, which are directly shown through the field changes $\Delta H_B$ of the B noises in §3.2 and the domain break down aberrances in §3.5. The calculated $\Delta H_B$ nicely fit with the experimental data.[9]

These numerical simulations in §3 clearly show the general magnetization characteristics such as the magnetization curves, the B effects and the $\Delta H_B$ distributions, which also coincide with the experimental data in good agreements.[9-13]

## 2. Magnetic Dipole Moment Interaction and Domain Energy

*2.1 Classical formulations*

The B effects are investigated variously in various materials.[14)-19)] The domain structures are also investigated in various materials using various methods.[20)-24)] For these many results, most of them are directly explained with the simulations using the domain energy calculations, which are based on the atomic dipole moment interactions in the classical theory. In this paper, the atomic dipole moments in Fe systems are set to be $n_b\mu_B$ composed of the Bohr magnetons $\mu_B$ with the effective spin number $n_b$ in an atom determined with the experimental data.

In classical theories, B-H characteristics are calculated with long range interactions between these magnetic dipole moments $\boldsymbol{\mu}_i$ and $\boldsymbol{\mu}_j$ localized at sites $i$ and $j$ respectively. The Bohr magneton: $\mu_B = eh/4\pi m = 9.274 \times 10^{-24}$ [Am$^2$] becomes the element of these magnetic dipole moments, where constants are $e=1.602 \times 10^{16}$ [C], $h=6.626 \times 10^{-34}$ [J.s] and $m=9.109 \times 10^{-31}$ [kg]. The atomic magnetic moment is replaced by a dipole moment of a magnetic rod with small distance vector $\boldsymbol{\delta}$ [m] and flux $\Delta\varphi$ or $\Delta\psi$ as

$$\boldsymbol{\mu}_B^0 = \mu_0\boldsymbol{\mu}_B = \boldsymbol{\delta}\Delta\varphi, \quad \boldsymbol{\mu}_i = n_b\boldsymbol{\mu}_B^0 = \boldsymbol{\delta}_i\Delta\psi \quad [\text{J.m/A}], \qquad (1)$$

where $\mu_0 = 4\pi \times 10^{-7}$ [H/m] is the vacuum magnetic permeability. In the body center cubic lattice (Fe), the regular dipole moment directions are drawn as like A B C in Fig. 1. In these types, the type A has the largest energy factors.[25)] In high temperature circumstances, these dipole moment directions distribute variously in thermal fluctuations. Such conditions explain that the domain structures do not depend on grain boundaries composed of various adjacent crystal directions. Under the restriction to be the type A, the structures of the dipole moment interactions in Fe are basically taken into two types of parallel and cross directions as shown in Fig. 2.

Fig. 1.

Fig. 2

Setting the distance vector $\boldsymbol{d}_{ij} = \boldsymbol{e}_{ij}d_{ij}$ between the dipole moments at $i$ and $j$, the moment interaction energies are equated using the Taylor expansion of $(d \pm \delta)^{-1/2}$ as

$$W_{ij} = \frac{1}{4\pi\mu_0 d_{ij}^3}\{(\boldsymbol{\mu}_i \cdot \boldsymbol{\mu}_j) - 3(\boldsymbol{\mu}_i \cdot \boldsymbol{e}_{ij})(\boldsymbol{\mu}_j \cdot \boldsymbol{e}_{ij})\}. \qquad (2)$$

Crystal Fe takes the BCC structure with the lattice constant $a=2.86\times10^{-10}$ [m] up to 911 ℃ and have 2 atoms in a lattice. This Fe metal has the dipole moments of $2n_b\mu_B^0$ per a lattice. Now, the distance $d_{ij}$ is represented using coefficients $c_{ij}$ and the constant $a$ as

$$d_{ij} = ac_{ij}. \qquad (3)$$

As for the parallel moment $\boldsymbol{\mu}_i$ and $\boldsymbol{\mu}_j$, the interaction energy $W_{ij}$ becomes

$$W_{ij} = \mu_0 \frac{n_b^2\mu_B^2}{4\pi d_{ij}^3}(1-3\cos^2\theta_{ij}) = E_a f_{ij}, \quad \delta<<d, \qquad (4)$$

$$E_a = \mu_0 \frac{n_b^2 \mu_B^2}{4\pi a^3} \to 1.812 \times 10^{-24} \quad \text{J}: \quad \text{Fe}(n_b = 2.22), \tag{5}$$

$$f_{ij} = (1 - 3\cos^2\theta_{ij})/c_{ij}^3, \tag{6}$$

where $\theta_{ij}$ are the angle between the distance vector $\mathbf{d}_{ji}$ and the moment vector $\boldsymbol{\mu}_i$. The position factors $f_{ij}$ are expanded to 4 terms in the $i$ $j$ lattice point representation with 2 atoms in a cubic lattice. The total energy $W_d$ [J] of the whole magnetic dipole moment interactions in a system become

$$f_j = \sum_i f_{ij}, \quad f = \sum_{i<j} f_{ij}, \tag{7}$$

$$W_j = \sum_i W_{ij}, \quad W_f = \sum_{i<j} W_{ij}, \tag{8}$$

where the count $i<j$ means half of the count $i, j$. The atomic dipole moments $\boldsymbol{\mu}_i$ in a cluster make the magnetization field $\mathbf{M}$ [T] for the moment $\boldsymbol{\mu}_j$ as like

$$\mathbf{M}_j = -\frac{1}{2}\sum_i \frac{1}{4\pi d_{ij}^3}\{\boldsymbol{\mu}_i - 3(\boldsymbol{\mu}_i \cdot \mathbf{e}_{ij})\mathbf{e}_{ij}\}. \tag{9}$$

Thus, in the external field $\mathbf{H}$ [A/m] by currents $\mathbf{I}$ [A], a system energy $W$ is equated as

$$W = W_H + W_f = -\sum_j \boldsymbol{\mu}_j \cdot (\mathbf{H} + \mathbf{M}_j/\mu_0)$$
$$= -(\sum_j \boldsymbol{\mu}_j) \cdot \mathbf{H} + E_a f, \tag{10}$$

where the magnetization field $\mathbf{M}$ is composed of the exact magnetic dipole moment arrays. It is the key point that the factor $f$ depends on the shape of the system. The magnetization $M_z$ for the field $H_z$ is derived from the $z$ component of the total dipole moment summation as

$$M_z = (\sum_j \boldsymbol{\mu}_j)_z. \tag{11}$$

In the simulation processes, the external field $H$ is divided into $N$ points in linear sweeps as

$$H_n = H_0(\frac{2n}{N} - 1): 1 < n < N,$$

$$H_n = H_0(3 - \frac{2n}{N}): N < n < 2N. \tag{12}$$

In these field alternations, the drastic domain structure changes are observed as the Barkhausen transitions constructing the terraces and the jumps in the magnetization curves.

*2.2 Energy factors between two dipole moments*

Eqs. (1)-(7) represent the interaction energy system of the dipole moments caused by the Bohr magneton. It is not proved that the dipole moment in Fe atoms has the same states of the Bohr magnetons of electron spins, but the calculated results with this condition show the good agreement with the experimental data. The quantum theory such as the Heisenberg model and Ising models cannot treat the long range interaction energies in (2) because of a large number of spins: $2^N$. Thus the exact calculations of the domain energies using the long range interactions are only discussed by classical ways, and there are not so many theoretical works in this subject.[1-5] Fortunately in Fe, the spin states affect as the magnetic dipole moments living in the quantum eigenstates of electrons. The magnetization energies are equated using such dipole moment interactions instead of the quantum spin interactions. In some magnetism, local spins interact through electrons in a conduction band, which do not behave as the dipole moments. In such case, the magnetizations are investigated with various approaches such as band theories.[26]

Here, the stable energy states are estimated using equations (1)-(10) in a body center cubic lattice of Fe. Setting the unit vector $e_i$ of the magnetic moment $\mu_i$ at the lattice point $i$ ;

$$e_i = e_x \mathbf{i} + e_y \mathbf{j} + e_z \mathbf{k} , \qquad (13)$$

the body center position vector is determined in each lattice point as

$$b = a(\mathbf{i}+\mathbf{j}+\mathbf{k})/2 .$$

Putting a $j$-th lattice point on the origin O, the energy factors of belt shape domains for the $j$-th moments at this origin are estimated in three dipole moment directions $e_A$, $e_B$, and $e_C$ of A, B and C type directions in Fig. 1:

$$\text{A: } e_i = \mathbf{k}, \quad \text{B: } e_i = (\mathbf{j}+\mathbf{k})/\sqrt{2}, \quad \text{C: } e_i = (\mathbf{i}+\mathbf{j}+\mathbf{k})/\sqrt{3}. \qquad (14)$$

In the parallel state, these directions counted by $i$ and $j$ are set to the same as

$$e_i = e_j, \qquad (15.\text{a})$$

and in cross moment states as

$$\text{A: } e_j = \mathbf{i}, \quad \text{B: } e_j = \mathbf{i}, \quad \text{C: } e_j = (-\mathbf{i}+\mathbf{j})/\sqrt{2} \qquad (15.\text{b})$$

in a domain. According to Fig. 2 and eq. (2) and setting $b=1/2$, the position vectors are defined for two sites $A_i$ and $a_i$ at lattice point $i$, and two sites $B_j$ and $b_j$ at lattice point $j$ respectively ;

$$r_{BA}(k,l,m) = r_{ba} = a(k\mathbf{i}+l\mathbf{j}+m\mathbf{k}) ,$$

$$c_{BA}(k,l,m) = c_{ba} = \sqrt{k^2+l^2+m^2} , \qquad (16)$$

$$r_{Ba}(k,l,m) = a\{(k+b)\mathbf{i}+(l+b)\mathbf{j}+(m+b)\mathbf{k}\} ,$$

$$c_{Ba}(k,l,m) = \sqrt{(k+b)^2+(l+b)^2+(m+b)^2} \qquad (17)$$

$$r_{bA}(k,l,m) = a\{(k-b)\mathbf{i}+(l-b)\mathbf{j}+(m-b)\mathbf{k}\} ,$$

$$c_{bA}(k,l,m) = \sqrt{(k-b)^2+(l-b)^2+(m-b)^2} . \qquad (18)$$

The angle relations between the above 2 dipole moments become

$$\cos\theta_{\alpha\beta}(k,l,m) = (r_{\alpha\beta} \cdot e_i)/r_{\alpha\beta} , \qquad (19.\text{a})$$

$$\sin\theta_{\alpha\beta}(k,l,m) = (r_{\alpha\beta} \cdot e_j)/r_{\alpha\beta} . \qquad (19.\text{b})$$

These 4 position factors between lattice points $i$ and $j$ in parallel direction are estimated as

$$f_{klm}^{F,AF} = 2\frac{1-3\cos^2\theta_{BA}}{c_{BA}^3} \pm \frac{1-3\cos^2\theta_{Ba}}{c_{Ba}^3} \pm \frac{1-3\cos^2\theta_{bA}}{c_{bA}^3} , \qquad (20.\text{a})$$

where the minus sign of the second and third terms indicates Anti-Ferro state. For the cross direction dipole moments, the 4 position factors are estimated as

$$f_{klm}^{F,AF} = -3(\frac{2\sin\theta_{BA}\cos\theta_{BA}}{c_{BA}^3} \pm \frac{\sin\theta_{Ba}\cos\theta_{Ba}}{c_{Ba}^3} \pm \frac{\sin\theta_{bA}\cos\theta_{bA}}{c_{bA}^3}) . \qquad (20.\text{b})$$

Using these formulas, full energy calculations between dipole moments are performed in treated systems. The calculated results using (13)-(15) show that the most stable dipole moment direction is the type A.

## 3. Simulations in nano-scale systems
*3.1 System representation*

We can simulate the Fe magnetization in nano-scale systems using the eqs. (9)-(21). Characters of nano-scale magnetizations are observed directly in nano-scale systems as shown in Fig. 3.

Fig. 3.

The system size is set in the number of lattice points as $N_x$, $N_y$ and $N_z$. Setting the values $e_1 = 1$, $e_2 = 1/\sqrt{2}$ and $e_3 = 1/\sqrt{3}$, the 26 dipole moment directions are represented using the marks as '>': $(e_1,0,0)$, '+': $(0,e_1,0)$, '1': $(0,0,e_1)$, '-': $(-e_1,0,0)$, '•': $(0,-e_1,0)$, 'W': $(0,0,-e_1)$, 'g': $(e_2,e_2,0)$, 'h': $(0,e_2,e_2)$, '^': $(e_2,0,e_2)$, 'j': $(-e_2,e_2,0)$, 'k': $(0,-e_2,e_2)$, '`': $(-e_2,0,e_2)$, 'm': $(e_2,-e_2,0)$, 'n': $(0,e_2,-e_2)$, ']': $(e_2,0,-e_2)$, 'p': $(-e_2,-e_2,0)$, 'q': $(0,-e_2,-e_2)$, '(': $(-e_2,0,-e_2)$, 's': $(e_3,e_3,e_3)$, 't': $(-e_3,e_3,e_3)$, 'u': $(e_3,-e_3,e_3)$, 'v': $(e_3,e_3,-e_3)$, 'w': $(-e_3,-e_3,e_3)$, 'x': $(-e_3,e_3,-e_3)$, 'y': $(e_3,-e_3,-e_3)$, 'z': $(-e_3,-e_3,-e_3)$. The main marks are shown in Fig. 3 (C). These arrays in the sheets show the domain patterns. The first domain sate is made of a down direction uniform array. Cooling are executed by *X-Z* plane traces of sites *j* from the front surface to the back surface with taking the energy minimum state of $W_j$ in (10) under full summations of the other sites *i*. The cooling trace for a *j* moment is executed by *c* time iterations under the condition as

$$(W_{j,c-1} - W_{j,c})/W_{j,c} < 10^{-4} \quad \text{or} \quad c < 40. \tag{21}$$

The steady states should represent low temperature states to be $T < 10$ [K]. Such calculations require the large scale computing resources in recent years. Using this simple orthorhombic lattice, many characteristics of the magnetizations are represented by changing the sizes, where near neighbor interactions become meaningless.

After next section, the experimental data are related to the simulations using the unit transformation as the field intensity $\mu_0 H = 10^{-4}$ T corresponding to $H$= 1 Oe = 79.577 A/m. The scale of the magnetization $M$ is normalized to be the atomic magnetic moment value $n_b$. The minimum energy states in (10) are determined after the annealing processes in (21).

### 3.2 Nano-thin film system and the Barkhausen effects

Here, a Fe nano-film of $N_x N_y N_z$=30×4×30 lattice points is simulated for clearing the Barkhausen effects and the results nicely coincide with the experimental data. The magnetization curve using the liner field change traces in (12) with $H_0$=2×10$^5$ A/m and $N$=150 steps ($\Delta H_k$=3×10$^3$ A/m) are represented in Fig. 4. The detail trace with $N$=500 steps of $\Delta H_k$=40 A/m width is shown in Fig. 5 to clear the B effects composed of the jumps and the terraces.

Fig. 4

The hysteresis structures of the magnetization curves are variously obtained in many references under various conditions. In these, the coercivities roughly become $H_c$=3.5~4.5×10$^4$ A/m =440~570 Oe being $\mu_0 H_c$=0.044~0.057 T in general nano-scale Fe magnetizations.[10)27)28)] In this square film, the coercivity $H_c$=3.83×10$^4$ A/m =480 Oe (0.048 T) and the remanent magnetization $B_r$/atom=1.27 normalized to $n_b$ adapted the observed value $n_b$=2.22 in bulk systems. Using the several layer systems of Fe nanoparticles on Al$_2$O$_3$/NiAl(100) [11)], the $<n_b>$ values are determined by the interactions between 3d and 4s orbitals composing chemical and metallic bond states including thermal fluctuations at temperature $T$. The value of $<n_b>$=2.22 should be generally adapted to large scale systems as discussed in ref. 11).

The precise magnetization curve in Fig. 5 shows the terraces and jumps in the B effects, where the major width becomes $\mu_0 \Delta H_a$=0.8×10$^{-3}$ T. As for the experimental data in ref. 29), the coercivities becomes $H_c$ =30 Oe at 10 K in something large Fe-film systems of 20~500 μm spot and 90 nm thick. In these data, the distribution $\rho(\Delta H)$ (peak is normalized to 1) for the terrace width $\Delta H$ are observed as $\rho$=1.0, 0.5, 0.25 and 0.2 for $\Delta H$=0.5, 1.0, 1.5 and 2.0 Oe respectively. This width $\Delta H_p$ at the density peak has the relation as $\Delta H_p/H_c$=0.5/30=1/60. To say, this result is same order of our result in Fig. 5 as $\Delta H_a/H_c$ =0.8×10$^{-3}$/0.048=1/60.

The domain structures in X-Z plane at a, b and c in Fig. 4 are drawn schematically in Fig. 6 and directly shown in Fig. 7 (a), (b) and (c) respectively. The figures (1) ~ (4) correspond to 4 sheets in the Y coordinate. The flux loop {right-down-left-up} constructions are clearly drawn with the continued marks such as { >> WW—11 }.

Fig. 6
Fig. 7

The schematic drawings in Fig. 6 show the flux vectors in the domains. The domains of cross square structures generally appear in the field alternations of **H**. This nature depends on the large minus structure factors of cross direction moments in (20.b). The down vectors out of the loops in (a) indicate the remanent magnetization flux of $B_r$. The fluxes constructing the two looped domains in (b) have no flow out flux. The up vector fluxes on the both side in (c) flow out from the system, where the double looped domains are observed similarly to (a) and (b). The domain structures take more sharp structures in lager systems.[9)] It is clearly observed in these figures that the B noises are caused by the break downs of the locally looped dipole moment constructions.

### 3.3 Nano-cube system

The magnetization in (11) under the alternate field with $\mu_0 H_0$=3×10$^5$ [T] in (12) are calculated in a nano-cube system of $N_x N_y N_z$=13×13×13 lattice points, where $H_c$=3.6×10$^4$ A/m (0.0452 T) and $B_r$/atom=0.34 ($n_b$=2.22) are obtained. The character of slow-saturated hysteresis curves appears at low temperatures as shown in Fig. 8. This reason is considered as that the interaction energies of dipole moments in cubic systems become 0 in a uniform direction and the energy minimum condition requires anti-direction moment arrays in nano-scale subsystems. This condition produces the slow-saturate magnetization for the external fields under local strong anti-Ferro domain structures. The domain structures at *a*, *b* and *c* points in Fig. 8 are represented in Fig. 9 about the 5 sheets in Fig. 3 (b).

Fig. 8

Fig. 9

The slow-saturate curves are observed experimentally in nanowire magnetizations encapsulated in aligned carbon nanotubes being non-saturate at room temperature[10] and in Fe layer systems individually composed of Fe nanoparticles saturated in 3kOe at 10 K [27)28)].

### 3.3 Nano-belt system

The magnetization curves in a nano-belt system of $N_xN_yN_z$=16×4×32 lattice points are calculated as in Fig. 10 with $H_c$=3.6×10$^4$ A/m and $B_r$/atom=1.88. These domain structures at *a*, *b* and *c* points are represented in Fig. 11 about every *X-Z* sheet.

Fig. 10

Fig. 11

Nanoparticle clusters joined with common fluxes show the strong $H_c$ and the high $B_r$ at zero temperature, [27)28)] which magnetization curves should correspond to the data in Fig. 10. The magnetization characters change in nanoparticle assemble systems as reported using Fe: $Al_2O_3$ nano-composite films.[12] In these large systems as > 250×250 nm$^2$, the $H_c$ has the tendency to be small according with the short joined distances and the massive particle densities as like 1/10 values comparing with the nano-particle cluster systems[27].

### 3.4 Long nano-belt system and domain break down avalanches

The strong coercivity $H_c$ and high and high remanent magnetization $B_r$ appears in the long nano-belt, -rod and -wire systems. For clearly representing the domain break downs, the long nano-belt system of $N_xN_yN_z$=16×4×64 lattice points is represented in this section, where $H_c$=4.14×10$^4$ A/m (520 Oe) and $B_r$/atom=2.07 ($n_b$=2.22) are obtained. The hysteresis curve is close to the square type structure. The dipole moments are going to uniformly array with longitudinal directions in the long body domains as like the marks of {WWW} and {111} in Fig. 13. The magnetizations extend step by step according with the increased field energy of $\Delta H_k$ in (10) as the Barkhausen effect, where the field densities are retarded from the self-consistent condition waiting for the increased field. The domain break down avalanches are strongly induced near the $H_c$ in long body systems as shown in Fig. 13 (a), (b) and (c), where every step of the increased field is $\Delta H_k = 4\times 10^3$ A/m.

Fig. 12

Fig. 13

The domains structures are not stable and distorted broken patterns are observed. These drastic transitions in the magnetizations are observed in various experiments[11)-22)]. The observation of negative Barkhausen jumps is reported in ref. 9) with permalloy thin-film microstructures as a violent case.

### 4. Summary

The magnetization mechanisms in Fe are directly represented by the computer simulations based on the domain energy calculations using the atomic magnetic dipole moment interactions under the classical theory. The results obtained by using large scale computing resources show the various nono-scale magnetization characteristics and nicely explain many experimental data. The nano-belt structure materials covered with non-magnetic materials could be easily created by spattering techniques with masks in plasma CVD. We can hope the appearance of the high ability magnets in these.

### Acknowledgements

This work is accomplished by the use of the supercomputer system in Institute of Solid State Physics, University of Tokyo. The author greatly thanks for the ISSP computer center.

**REFERENCS**

1) S.L.A. de Queiroz and M. Bahiana : Phys. Rev. E. **64**, (2001) 066127-1-6.
2) H.T. Savage, D-X. Chent, C. Go'mez-Polo, M. Va'zquez, and M. Wun-Fogle : J. Phys. D; Appl. Phys. **27** (1994) 681-684.
3) T. Koyama : Sci. Technol. Adv. Mater. **9** (2008) 013006_1-9.
4) H. Kronmüller, H.-R. Hilzinger, P. Monachesi, and A. Seeger : Appl. Phys. **18** (1979) 183-193.
5) S. Chikazumi: Physics of Ferromagnetism (2nd edn. Oxford: Oxford University Press, 1997) pp. 1-10.
6) B. Hillebrands, and K. Ounadjela: Spin dynamics in confined magnetic structures I, (Springer, 2002).
7) S. Obata : IEEJ Trans. FM, **131** (2011) 838-845.
8) S. Obata : J. Magn. Soc. Jpn., **36** (2012) 161-168.
9) Shuqiang Yang, G.S.D. Beach, and J.L. Erskine: J. Appl. Phys. **100** (2006) 113914
10) B.C. Satishkumar, A. Govindaraj, P.V. Vanitha, Arup K. Raychaudhuri, and C.N.R. Rao: Chem. Phys. Let. **362** (2002) 301-306
11) N.M. Dempsey, L. Ranno, D. Givord, J. Gonzalo, R. Sema, G.T. Fei, A.K. Petford-Long, R.C. Doole, and D.E. Hole: J. Appl. Phys, 90 (2001) 6268-6274.

12) Wen-Chin Lin, C.B. Wu, P.J. Hsu, H.Y. Yen, Zheng Gai, Lan Gao, Jian Shen, and Minn-Tsong Lin: J. Appl. Phys., **108** (2010) 034312.
13) S. Zapperi, and G. Durin : Comp. Mater. Science, **20** (2001) 436-442.
14) S. Zapperi, P. Cizeau, G. Durin, and E. Stanley : Phys. Rev. B, **58** (1998) 6353-6366.
15) F. Colaiori, and A. Moro : Advances in Physics, **57** (2008) 287-359.
16) S. Yang and J.L. Erskine : Phys. Rev. B, **72** (2005) 064433-1-13.
17) K. Kova'cs, and Z. Ne'da : J. Opt. Elect. Adv. Mater, **8** (2006) 1093-7.
18) D.A. Christian, K.S. Novoselov, and A.K. Geim : Phys. Rev. B, **74** (2006) 06443_1-6.
19) G. Durin, and S. Zapperi : J. Stat. Mech., **2006** (2006) P01002_1-11.
20) D.C. Jiles : Czech. J. Phys., **50** (2000) 893-988.
21) O. Gutfleisch, K.-H. Müller, K. Khlopkov, M. Wolf, A. Yan, R. Scha"fer, T. Gemming, and L. Schultz : Acta Materialia, **54** (2006) 997-1008.
22) T. Eimu"ller, P. Fischer, G. Schu"tz, P. Guttmann, G. Schmahl, K. Pruegl, and G. Bayreuther : J. Alloys and Compounds, **286** (1999) 20-25.
23) P. Fischer, M.-Y. Im, T. Eimu"ller, G. Schütz, and S.-C. Shin : J. Magn. Magn. Mater, **286** (2005) 311-314.
24) N.I. Vlasova, G.S. Kandaurova, and N.N. Shchegoleva : J. Magn. Magn. Mater, **222** (2000) 138-158.
25) S. Obata : IEEJ Trans. FM, **133** No 9 (2013) 489-499.
26) S. Wakoh, and J. Yamashita : J. Phys. Soc. Jpn., **21** (1966) 1712-1726.
27) M. Yoon, Y.M. Kim, Y. Kim, V. Volkov, H.J. Song, Y.J. Park, S.L. Vasilyak, and I.-W. Park: J. Magn. Magn. Mater. **265** (2001) 357-362.
28) K.D. Sorge, J.R. Thompson, T.C. Schulthess, F.A. Mondine, T.E. Haynes, S. Honda, A. Meldrum, J.D. Budai, C.W. White, and L.A. Boatner : IEEE Trans. Magn. **37** (2001) 2197-2199.
29) E. Puppin and M. Zani : J. Phys.: Condens. Matter **16** (2004) 1181-1188.

FIGURES
Fig. 1.

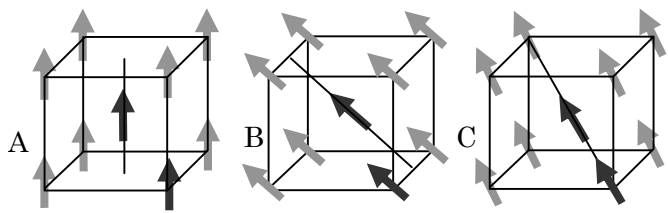

Fig. 1   Ferro magnetic moments in a BCC lattice. The dipole moment directions are mainly divided to 3 types of A, B and C. The domain energies have the largest value in the type A.

Fig.2

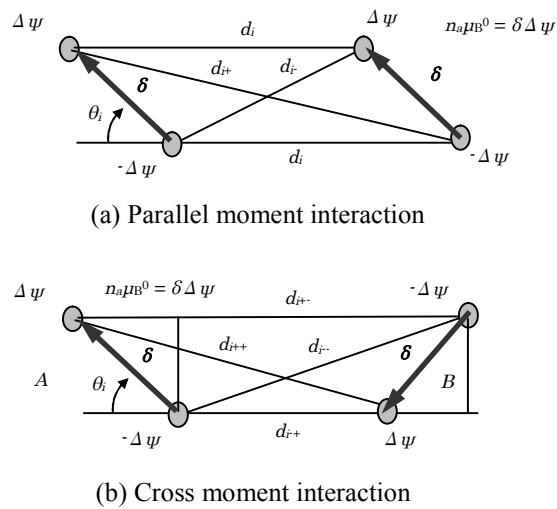

Fig. 2   Two type interactions between magnetic moments.

Fig.3

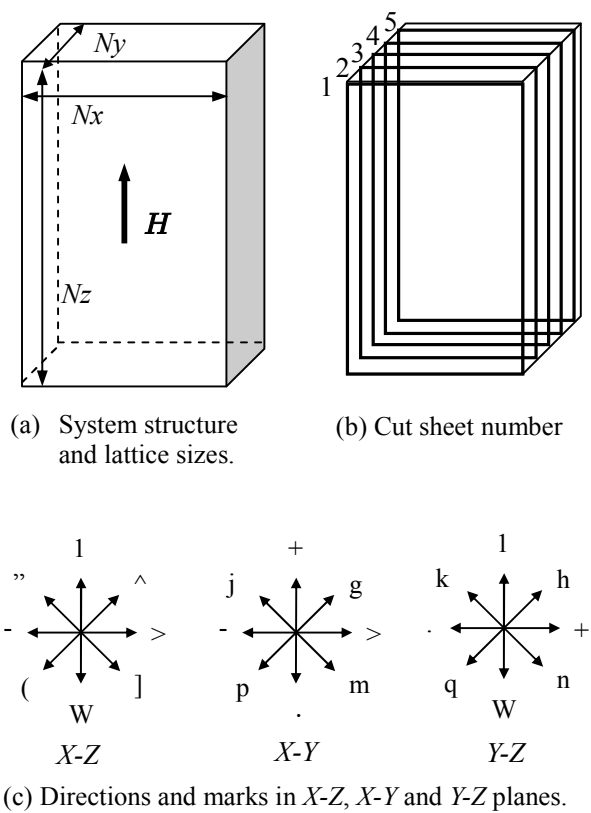

(a) System structure and lattice sizes.

(b) Cut sheet number

(c) Directions and marks in *X-Z*, *X-Y* and *Y-Z* planes.

Fig. 3. Nano-scale system size and divided planes in *Y* axis and direction marks. (a) $N_x$, $N_y$ and $N_z$ are the lattice point numbers of *X*, *Y* and *Z* axis in a orthorhombic system. (b) Domain structures in a system are represented in cut sheets. (c) Direction marks in *X-Z*, *X-Y* and *Y-Z* planes.

Fig. 4

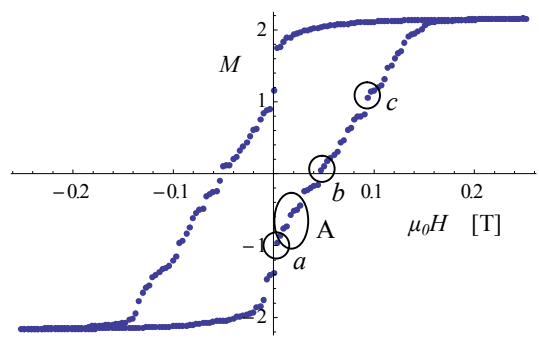

Fig. 4   Magnetization curve ($n_b$=2.22) in a Fe nano-film composed of $N_xN_yN_z$=30×4×30 lattice points with the field $H$=2×10$^5$ [A/m]. The area A is precisely traced in Fig. 5 for showing the B jumps and the terraces. The domain structures at the points *a*, *b* and *c* are drawn in Fig. 6.

Fig. 5

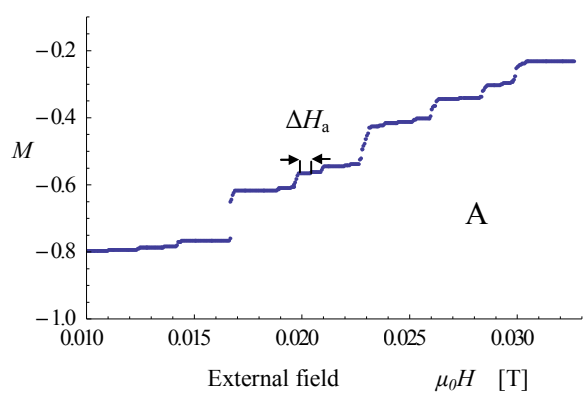

Fig. 5   The precise magnetization curve at the area A in Fig. 4. The representative widths of terraces are $\Delta H_a$=$H_c$/60.

Fig. 6

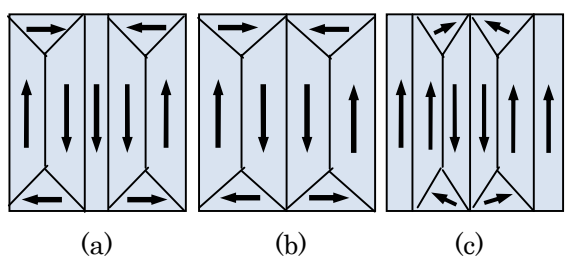

Fig. 6　Schematic domain structures at point *a*, *b* and *c* in Fig. 5 correspond to real simulated results (a), (b) and (c) in Fig. 7 respectively.

Fig. 7

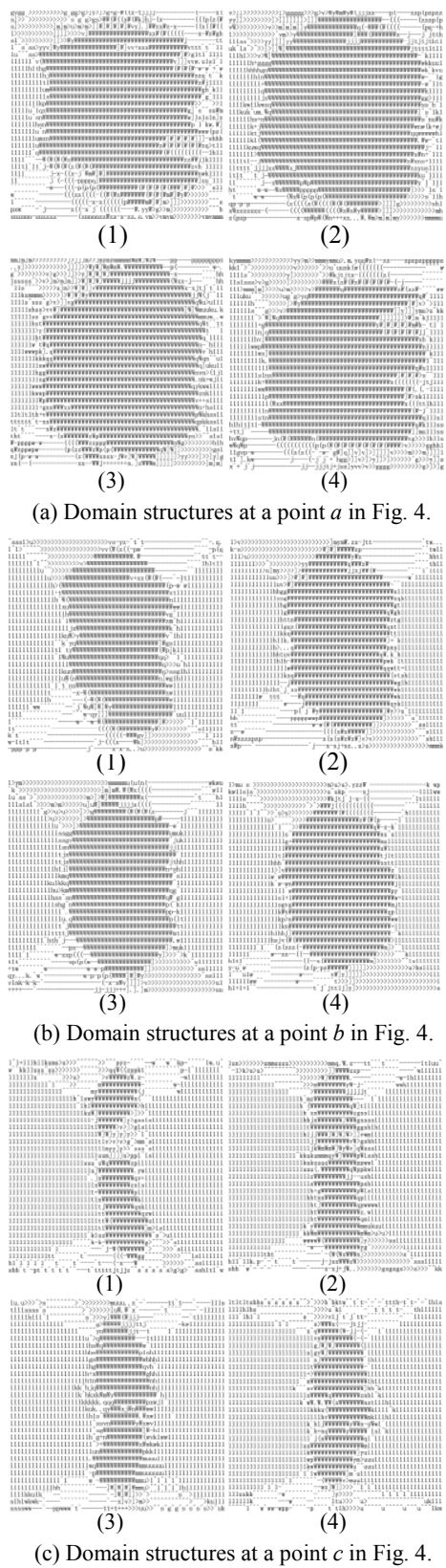

(1) (2)

(a) Domain structures at a point *a* in Fig. 4.

(1) (2)

(3) (4)

(b) Domain structures at a point *b* in Fig. 4.

(1) (2)

(3) (4)

(c) Domain structures at a point *c* in Fig. 4.

Fig. 7  The *X-Z* domain structures of 4 sheets in the Fe nano-film system of 30×4×30 lattice points, which correspond to the point *a*, *b* and *c* in Fig. 4. The schematic figures are drawn in Fig. 8

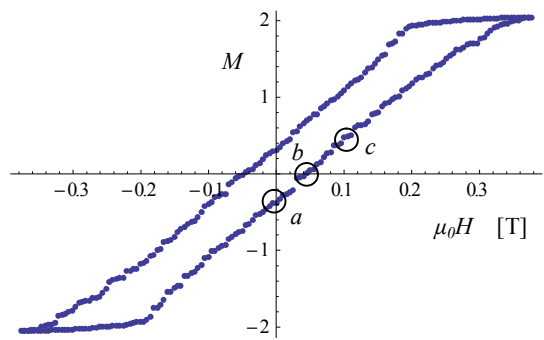

Fig.8 Magnetization curve in a Fe nano-cube system of 13×13×13 lattice points with the field $H_m$=3×10$^5$ [A/m]. The domain structures at the points *a*, *b* and *c* are drawn in Fig. 9.

Fig. 9.

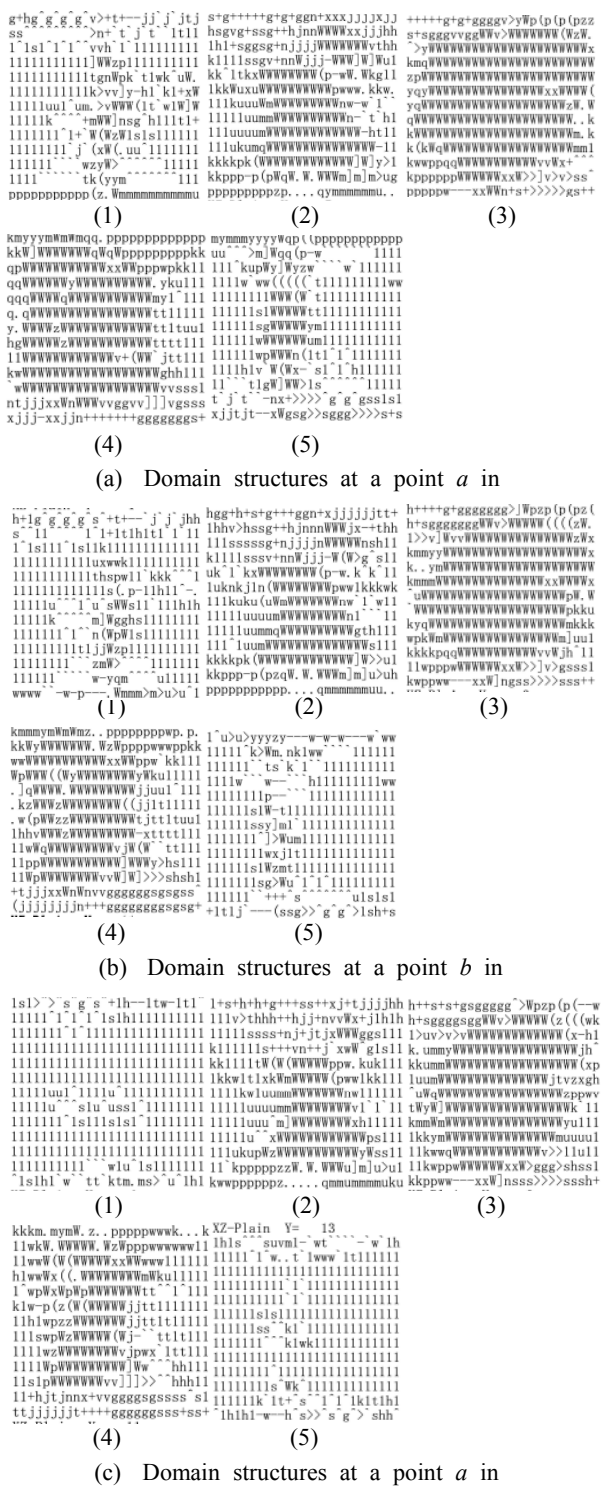

(a) Domain structures at a point *a* in

(b) Domain structures at a point *b* in

(c) Domain structures at a point *a* in

Fig. 9  The domain structures in the Fe nano-cube system of 13×13×13 lattice points, which correspond to the point *a*, *b* and *c* in Fig. 8. The numbers (1)-(5) represent the cut sheet 1, 4, 7, 10 and 13 as in Fig. 3 (b).

Fig. 10.

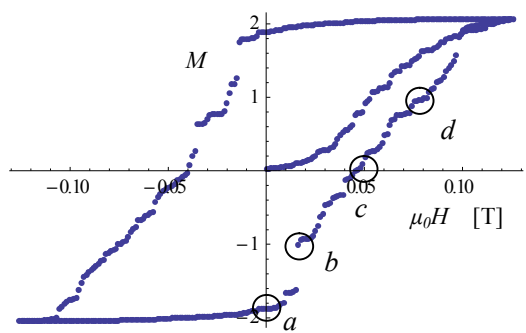

Fig. 10  Magnetization curve in a Fe nano-belt system of 16×4×32 lattice points with the field $H_m=1\times10^5$ [A/m]. The domain structures at the points *a*, *b* and *c* are drawn in Fig. 11. The typical patterns are observed.

Fig. 11.

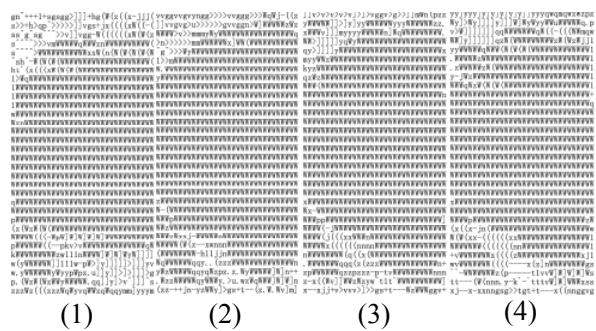

(1) (2) (3) (4)

(a) Domain structures at a point *a* in Fig. 9.

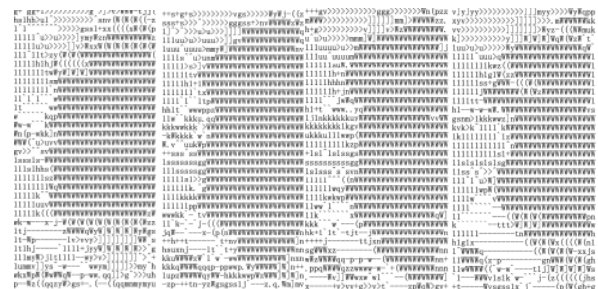

(1) (2) (3) (4)

(b) Domain structures at a point *b* in Fig. 9.

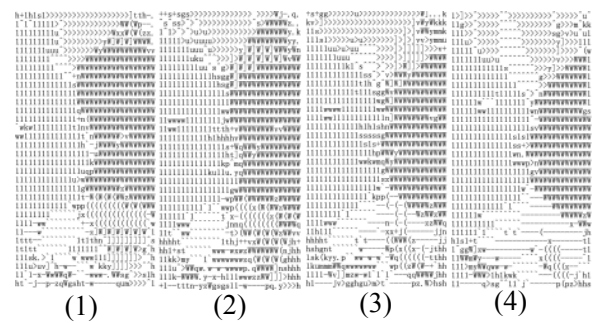

(1) (2) (3) (4)

(c) Domain structures at a point *c* in Fig. 9.

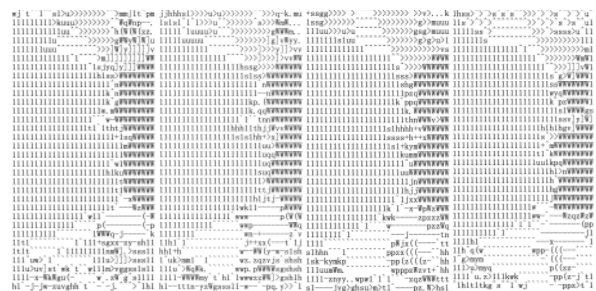

(1) (2) (3) (4)

(d) Domain structures at a point *d* in Fig. 9.

Fig. 11. The *X-Z* domain structures of 4 sheets in the Fe nano-belt 16×4×32 system. (a), (b), (c) and (d) correspond to the *a*, *b*, *c* and *d* points in Fig. 10 respectively.

Fig. 12.

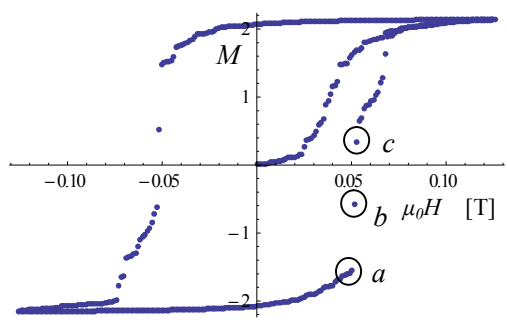

Fig. 12  Magnetization curve in a Fe nano-belt system of 16×4×64 lattice points with the field $H_m$=1.0×10$^5$ [A/m]. The domain structures at the points *a*, *b* and *c* are drawn in Fig. 13. The point *b* is a large scale avalanche state which needs the cooling processes over 40 times in $\Delta H$=10$^{-3}$ [A/m]. The hysteresis curve becomes the square type structure.

Fig. 13.

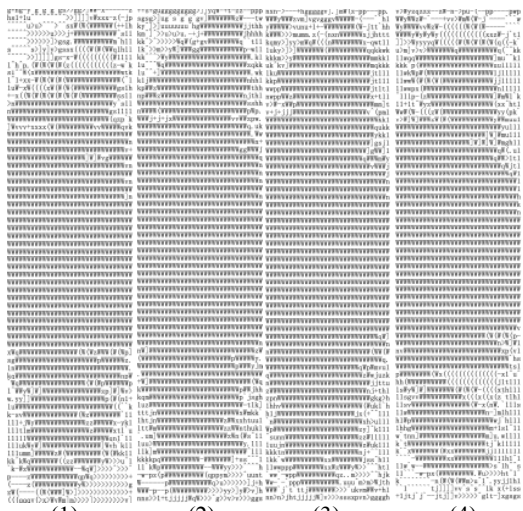

(1)　　　　(2)　　　　(3)　　　　(4)
(a) The domain structures at the point *a* in Fig. 11.

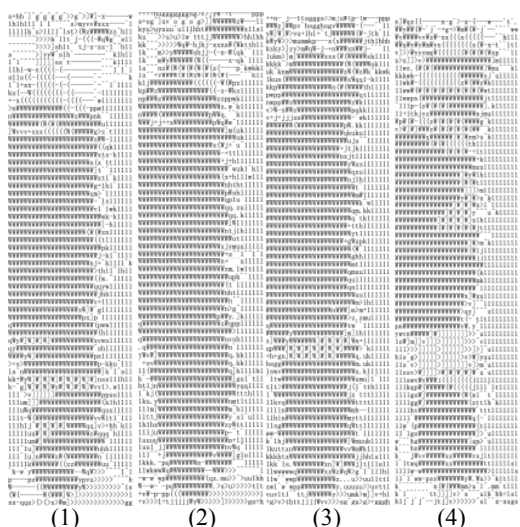

(1)　　　　(2)　　　　(3)　　　　(4)
(b) The domain structures at the point *a* in Fig. 11.

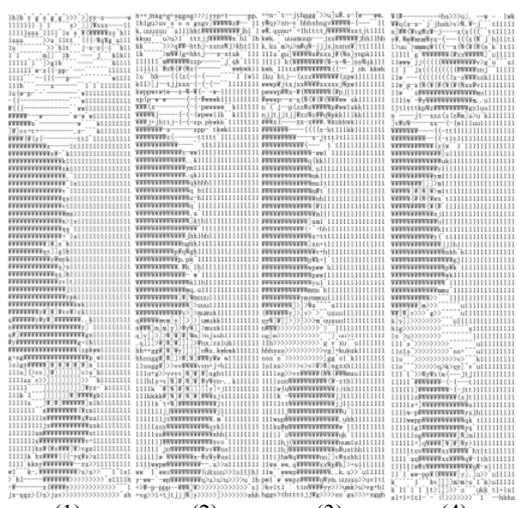

(1)　　　　(2)　　　　(3)　　　　(4)
(c) The domain structures at the point *a* in Fig. 11.

Fig. 13　The *X-Z* domain structures of 4 sheets in the Fe nano-belt 16×4×64 system. (a), (b) and (c) correspond to the *a*, *b* and *c* points of the avalanche states in Fig. 12 respectively.